\def\mearth{{\rm\,M_\oplus}}
\def\bbc{{\beta / \beta_\mathrm{crit}}}
\def\gsim{~\rlap{$>$}{\lower 1.0ex\hbox{$\sim$}}}
\def\lsim{~\rlap{$<$}{\lower 1.0ex\hbox{$\sim$}}}
\def\etal{{\it et al.\thinspace}}

\documentclass[12pt, preprint]{aastex}
\usepackage{graphicx}
\usepackage{natbib,color,lscape}

\title{Stability of additional planets in and around the habitable zone of
 the HD 47186 Planetary System}
\author{Ravi kumar Kopparapu\altaffilmark{1,4}, Sean N. Raymond\altaffilmark{2,4}, 
Rory Barnes\altaffilmark{3,4}}

\altaffiltext{1}{Center for Gravitational Wave Physics, 104 Davey lab,
Pennsylvania State University, University Park, PA - 16802-6300, USA;
ravi@gravity.psu.edu}
\altaffiltext{2}{Center for Astrophysics and Space Astronomy, University of
 Colorado, UCB 389, Boulder CO 80309- 0389}
\altaffiltext{3}{Department of Astronomy, University of Washington, Seattle,
WA, 98195-1580}
\altaffiltext{4}{Virtual Planetary Laboratory}

\begin{document}
\begin{abstract}

We study the dynamical stability of an additional, potentially habitable
planet in the HD 47186 planetary system. Two planets are currently
known in this system: a ``hot Neptune'' with a period of $4.08$ days and a Saturn-mass planet
with a period of $3.7$ years. Here we consider the possibility that one or more undetected planets exist between the
two known planets and possibly within the habitable zone in this
system. Given the relatively low masses of the known planets,
additional planets could have masses $\lsim 10 \mearth$, and hence be
terrestrial-like and further improving potential habitability. We
perform $N$-body simulations to identify the stable zone between planets $b$
and $c$ and find that much of the inner habitable zone can harbor a $10 \mearth$ planet. With the current radial
velocity threshold of $\sim 1$ m/s, an additional planet should be 
detectable if it lies at the inner edge of the habitable zone at 0.8 AU.  We
also show that the stable zone could contain two additional planets of $10
\mearth$ each if their eccentricities are lower than $\sim 0.3$.

\end{abstract}

\section{Introduction}
\label{sec1}

Recent advances have brought the astronomical community to the verge of
discovering a terrestrial planet in the habitable zone (HZ) of its parent star.
Classically, the HZ is defined as the range of orbits for which a
terrestrial mass planet ($0.3 \lesssim M \lesssim 10 \mearth$), with favorable
atmospheric conditions, can sustain liquid water on its surface
\citep{Kasting1993, Selsis2007}. Many giant planets have been discovered in
the HZ, and low-mass planets have been found very close to the
HZ \citep[see][for example]{Udry2007, Selsis2007, vonBloh2007}, but no serious candidates
for habitability have been found.

Recently \cite{Bouchyetal2008} used the HARPS spectrograph (Mayor et al 2003)
to detect two planets around the nearby star HD 47186: a hot Neptune (HD 47186 b; $22.78 ~
\mearth$) with an orbital period of $4.08$ days and a Saturn-mass planet (HD
47186 c; $0.35 ~ M_\mathrm{Jup}$) with a period of $ 3.7 $ years.  The orbital
elements for this system are given in Table \ref{table1}.  A curious
characteristic of the host star HD 47186 is that its mass and luminosity are
very similar to our own Sun, albeit with higher metallicity (Table 1. of
\cite{Bouchyetal2008}).  Therefore the HZ of this
system may be similar to our own solar system's, and hence lies
between the two known planets. \cite{Bouchyetal2008} found that the
combined radial velocity (RV) amplitude for this system is $4.30$ m
s$^{-1}$ with $0.91$ m s$^{-1}$ residuals and reduced $\chi^{2} =
2.25$. So potentially habitable planets may lie at or below the
current RV detection threshold but may be discovered by
further observations.  Here we perform $N$-body simulations to assess
the stability of additional hypothetical planets within the HZ of HD 47186. The relatively small masses of the two known planets
suggest an additional planet would likely also have low mass. Masses
less than $10 \mearth$ are often considered the upper limit for
terrestrial-like planets \citep{Pollack96,BHL00,Hubickyj05}, hence an
additional planet in the HZ of this system might have a mass
conducive to habitability.

As of Feb 27, 2009, 290 extra-solar planetary systems are known including 37
multiplanet systems\footnote{www.exoplanet.eu}. \cite{BQ04} found that many
multiplanet systems tend to be close to the edge of stability such that slight
perturbations to the system would result in destabilization \citep
{BQ01,Gozd01,GM01,Erdi04}. This observation led to the
``Packed Planetary systems (PPS)'' hypothesis (Barnes \& Raymond 2004;
Raymond \& Barnes 2005; Raymond \etal 2006; Barnes \& Greenberg 2007; see also Laskar 1996) which postulates that if a
region of stability is available in between two known planets, then that
region will harbor an unseen planet (in other words, planetary systems tend to
be ``packed'', with no large gaps between planets).  To constrain the location
of additional planets, a large number of studies have mapped out stable zones
in known extra-solar planetary systems, mainly using mass less test particles
(e.g., Rivera \& Lissauer 2000; Jones et al 2001; Menou \& Tabachnik 2003;
Dvorak et al 2003; Barnes \& Raymond 2004).  Recently, \cite{Bean2008}
reported the discovery of a Saturn-mass planet in the narrow,
well-characterized stable zone of HD 74156 \citep{BR2004, RB2005}.  This
constituted the first prediction of the orbit and mass of an
extra-solar planet (Barnes et al. 2008a)\footnote{It  should be
 noted that HD 74156 d remains a controversial detection;
 \cite{Wittenmyer2009} in a reanalysis of the \cite{Bean2008}
 spectra see no evidence for a third planet orbiting HD 74156.}.

The PPS model was further developed by the identification of an
analytic stability boundary in two-planet systems
\citep{BG06,BG07}. The boundary lies where the ratio
of two quantities, $\bbc$, approximately equals unity. Here $\beta$
depends on the total angular momentum and energy of the system and
$\beta_\mathrm{crit}$ depends only on the masses \citep{MB1982, Gladman1993}.
If $\bbc \gsim 1$ (and no resonances are present), then the system is
stable. Moreover, Barnes et al. (2008a) found that systems tend to be
packed if $\bbc \lesssim 1.5$ and not packed when
$\bbc \gtrsim 2$. \cite{BG07} pointed out that the vast majority of
two-planet systems are observed with $\bbc < 1.5$ and hence are packed.


Our investigation into possible habitable planets in HD 47186 is
further motivated by the large separation between the two known
planets: the $\bbc$ value for this
system is $6.13$, the largest value among known, non-controversial
systems that have not been affected by tides.\footnote{See
http://www.astro.washington.edu/users/rory/research/xsp/dynamics/ for
an up to date list of $\bbc$ values for the known extra-solar multiple
planet systems.}  This value is much higher than the packing limit,
giving further support to the argument that at least one additional
companion might exist between the two known planets. Moreover, the
success of $N$-body investigations in predicting the mass and orbit
of additional planetary companions suggests a similar analysis on HD
47186 could yield similar results. But this time, the
yet-to-be-detected planet could be habitable. In \S\ref{sec2} we
describe our numerical methods and map out stable regions for a
hypothetical $10 \mearth$ planet $d$ between the two known planets.
Following \cite{raymondetal2008}, we then investigate the
perturbations of this hypothetical planet on the orbits of the known
planets $b$ and $c$.  This approach allows us to exclude a portion of
parameter space that is dynamically stable but offers a low
probability of detection.  Finally, in
\S\ref{sec3} we discuss our results.

\section{Dynamical stable zones in HD 47186}
\label{sec2}

\subsection{Numerical Simulations}

We started our simulations by assigning the two known planets from HD 47186
their best-fit orbits and masses from Table 1 \citep{Bouchyetal2008},
and assuming mutual inclinations of 1 degree.
 We also included a hypothetical planet $d$ in between the
orbits of planets $b$ and $c$.  From simulation to simulation, we varied
planet $d$'s semi-major axis in the range $ 0.06 \le a_\mathrm{d} \le 2.32$ AU in
steps of $0.02$ AU and eccentricity in the range $ 0 \le e_\mathrm{d}
\le 0.6$ in steps of
$0.05$, for a total of 1483 simulations.  We assigned planet $d$ a mass of $10
~\mearth$, which would induce in the star a radial velocity amplitude of
 3.5-0.6
m s$^{-1}$ in the range that we considered ($0.06 - 2.3$ AU).
  We evolved each system with the {\tt Mercury}
\citep{Chambers1999} integrator, using the symplectic Wisdom-Holman mapping
(Wisdom \& Holman 1991).  Each system was integrated with a time step of 1.0
day and conserved energy to better than 1 part in $10^{5}$.  Simulations were
stopped after $10$ Myr or if there was a close encounter between any two
planets of less than $3$ Hill radii.  Previous work has shown that most
unstable planetary systems undergo instabilities on a timescale of 1 Myr or
less (e.g., Barnes \& Quinn 2004).  Thus, our 10 Myr integrations should
be sufficient to give a realistic depiction of the stable zone.

\subsection{Mapping stable zones}

In Fig.~\ref{fig1} we show the results of our simulations plotted in
$a$--$e$ space. The green and red filled circles represent stable and
unstable simulations, respectively. The two known planets in HD 47186
are shown as blue filled circles, and the vertical bars represent the
observational uncertainties in their
eccentricities. The two curves overlaying the stable/unstable points
are the crossing orbits for planets $b$ and $c$. The blue shaded
background is the ``eccentric habitable zone'' (EHZ) for this system
(Barnes et al 2008b).  The EHZ is defined as the HZ from Selsis et al
(2007), assuming $50 \%$ cloud cover, and modified to take into
account variations in the orbit-averaged flux with eccentricity, which
appears to be the key factor in determining surface temperature
(Williams \& Pollard 2002).  Also labeled are some of the mean-motion
resonances (MMRs) between the hypothetical planet $d$ and outer
massive planet $c$. 
 Out of
the 1483 simulations, 856 ($57 \%$) simulations were stable for 10
Myr, and the stable zone includes the inner half of the EHZ. Within
this stable zone, several MMRs produce ``islands'' of instability
associated with specific MMRs, notably $5d:1c$, $7d:2c$, and
$5d:2c$. Conversely, beyond the crossing orbit of planet c (black 
dashed  curve overlapping EHZ), there are several MMRs, such as $2d:1c$, $7d:3c$
(not shown, but lies in between $5d:2c$ and $2d:1c$) and $3d:2c$ that
provide stability.  This result is similar to that of Raymond et al
(2008), who found regions of stability/instability associated with
mean motion resonances in the 55 Cancri planetary system.

Using just dynamical stability as a metric, all of the stable (green) points
in Fig.~\ref{fig1} should be equally likely to support a $10 \mearth$ planet.  To
narrow down the possible orbit of planet $d$, we calculate a quantity called ``Fraction of
Time spent on Detected orbits'' (FTD, \cite{raymondetal2008}), which takes into
account the back-reaction of planet $d$ on the orbits of planets $b$ and $c$.
The FTD provides a measure of the probability of finding planets $b$ and $c$
on their current best-fit orbits, including observational uncertainties (see
Table 1), taking into account perturbations from planet $d$.  In practice, the
FTD simply evaluates the time spent by each observed planet ($b$ and $c$) on
their best-fit orbit, to within the uncertainties.  A stable hypothetical
planet can cause oscillations in the known planets' orbits that may be
larger than the uncertainties.  If the FTD value is
small, then the oscillation amplitudes are large and the
chance of observing the system in it's current configuration is
small. Hence, it is less likely for a
hypothetical  planet $d$ to have those orbital parameters.  If the FTD value is
close to 1, then it is more probable for planet $d$ to be in that
orbit. Note, however, that some systems may be observed in unlikely
configurations (Veras \& Ford 2008).

Figure~\ref{fig2} shows the FTD for the stable region mapped out in
Fig.~\ref{fig1}: yellow is the most likely region for planet $d$ to
exist (high FTD), and black is the least likely (low FTD; see color
bar).  Comparing Figs.~\ref{fig1} $\&$ \ref{fig2}, the region $a <
0.2$ AU is an unlikely location for planet $d$ to exist.  On the other
hand, beyond $a \approx 0.3$ AU, the FTD value is uniformly close to
1. Therefore planet $d$ does not significantly perturb planet $c$
(otherwise there would have been low FTD values), and acts effectively
as a massless particle.  This leaves a large contiguous stable region
with high FTD values from $0.3 - 1.3$ AU.
We see that planet $d$ might exist in the habitable zone of HD 47186.  52.5\%
of the stable simulations are within the EHZ, but it is not clear how
likely it is that planet $d$ should exist there.

We also performed $30$ simulations to see if two additional planets
could exist in the stable zone.  We placed two planets of $10 \mearth$
each in the range $ 0.3 \le a_\mathrm{d} \le 1.3$ AU and $0 \le
e_\mathrm{d} \le 0.6$.  We found that $16$ cases resulted in unstable
simulations and $14$ stable. Most of the stable simulations ($13, \sim
93 \%$) had low eccentricity ($e < 0.3$). In some stable
cases, both hypothetical planets were in the EHZ.

Additional simulations assuming that the masses of the known
planets b \& c are twice their minimum mass, equivalent to an
inclination angle of 30 degrees,
did not show any significant change in the stability boundaries.
 We found that the inner stability boundary near planet b is not at all
 affected, whereas the region between 1.5-1.8 AU is slightly modified.
 Specifically, regions stabilized by 2:1 mean motion resonance near 1.5 AU,
 are now smaller with some of the stable points turning into unstable.
 Overall, there is no significant change in the stability of the HZ.

\section{Discussion}
\label{sec3}

\begin{figure}[!hbp|t]
\includegraphics[width=\textwidth]{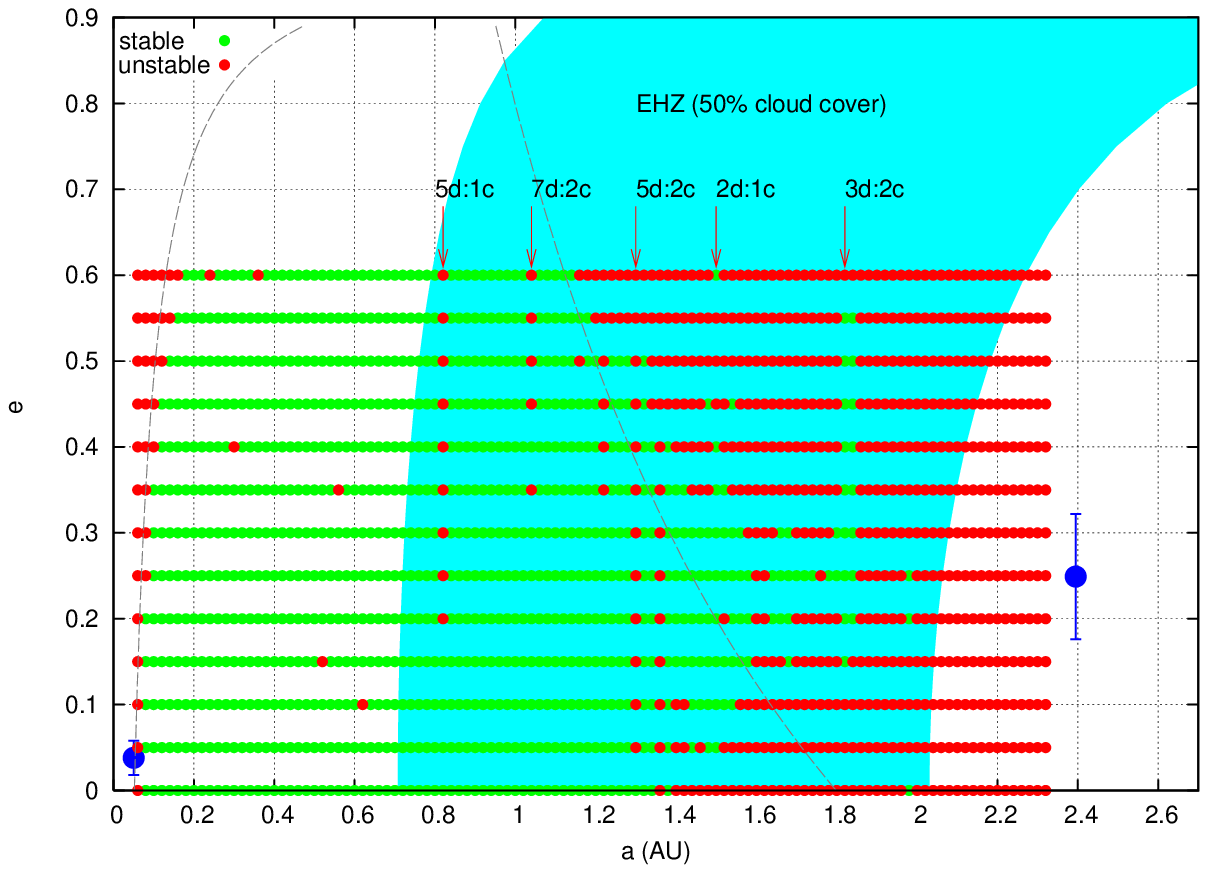}
\caption{The stable zone for a hypothetical $10 \mearth$ in between the two
known planets of HD 47186. The red and green circles represent stable and
unstable simulations, respectively. The known planets HD 47186 $b$ and $c$ are
indicated by blue filled circles: the error bars represent the observational
uncertainties in the best-fit eccentricities. The curves on top of the
stable/unstable zones represents crossing orbits for planets $b$ and $c$. The
blue background shows the eccentric habitable zone \citep{Barnes2008} for this
system.  Also shown are the several MMRs between planet $d$ and $c$.}
\label{fig1}
\end{figure}

\begin{figure}[!hbp|t]
\includegraphics[width=\textwidth, angle=0]{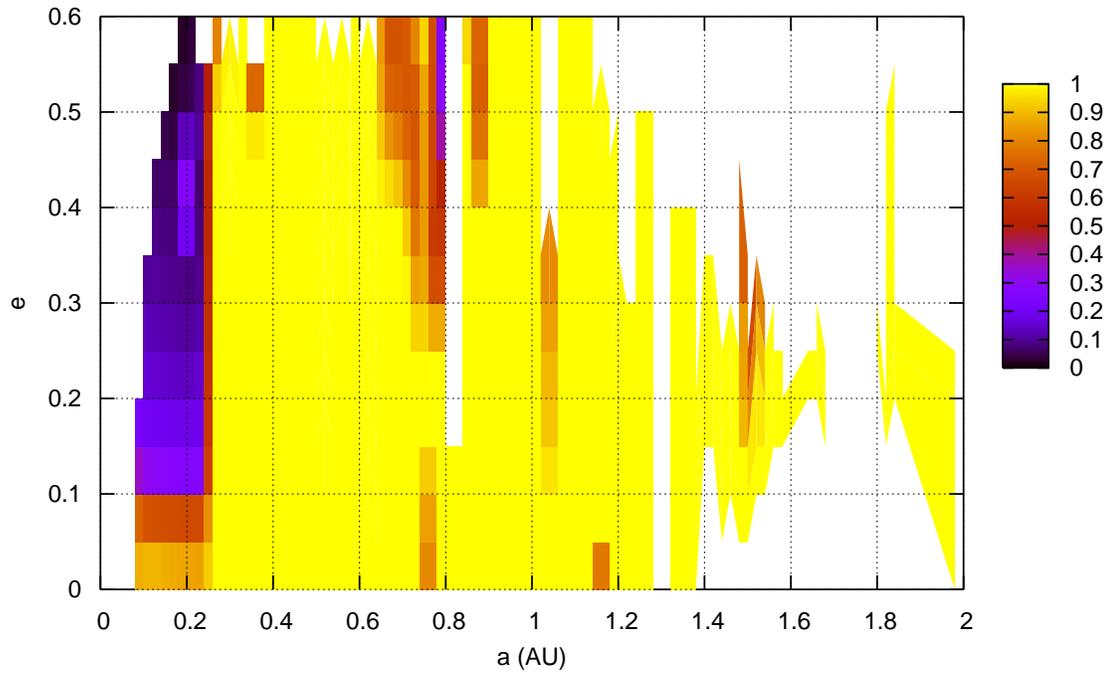}
\caption{The FTD (``Fraction of Time spent on Detected orbits'')
(\citep{raymondetal2008}) for the stable zone of HD 47186. The FTD value is
shown in the color bar: yellow indicates a high likelihood of detecting planet
$d$ with those orbital parameters. The long gaps correspond to unstable zones
arising from MMRs. }
\label{fig2}
\end{figure}

We have shown that a hypothetical
$10 \mearth$ planet can exist between planets HD 47186 $b$ and
$c$ for orbital distances of roughly 0.1 -- 1.3 AU (Fig. \ref{fig1}). Our
FTD analysis shows that this planet is more likely to be found
between $\sim 0.3$ and $1.3$ AU (Fig. \ref{fig2}). We also find that
two $10 \mearth$ planets with low eccentricities can exist between
planets b and c. A $10 \mearth$ planet between  0.3 and 0.7 AU (the high FTD region
in Fig. \ref{fig2}) produces an RV amplitude of $1 - 2$ m/s, and should be detectable in the near future.  It is possible that the two known planets in 
HD 47186 may have migrated to their present location. 
Previous studies \citep{rms2006, mandell2007}
 have shown that migrating giant planets don't preclude the existence
of terrestrial planets. In fact,  if the outer
planet migrated inward then the most likely place for an
extra planet would be just interior to the 2:1 MMR, which lies in the middle
of the EHZ.  Planets of $\sim 10
\mearth$ or less are thought to be the largest potentially terrestrial planets
with rocky surfaces and thin atmospheres.  This provides an effective
upper limit for habitability as we know it. Therefore HD 47186 offers
the chance to detect a terrestrial planet in the HZ of a nearby,
solar-type star.
 Roughly 30\% of systems are thought to harbor
super-Earth sized planets ($M_p \lesssim 10 \mearth$, \citep{Mayor2009}).
 If this
frequency increases for lower-mass planets,
the (non-)detection of HD 47186 d could shed light on the fraction of
stars with terrestrial-like planets in the habitable zone, sometimes called
$\eta_\oplus$.  If, in a complex system such
as this one, a terrestrial-scale planet can form and survive, then perhaps
$\eta_\oplus$ is relatively large, increasing the likelihood the spacraft
such as {\textit {Kepler}} and {\textit {Terrestrial Planet Finder}} will find
habitable planets.
 We encourage observers to use our results to assist
in the search for HD 47186 d.

The HD 47186 system is also another critical test of PPS hypothesis,
which asserts that planetary systems should not contain large empty
gaps. 
The PPS hypothesis follows naturally from the planet-planet scattering
model \citep{retal2009} and clearly predicts that there exists an
 additional planet in the gap between planets $b$ and $c$.
 In fact, two $10
\mearth$ planets could exist in the stable zone if they are at low
eccentricity.  The connection between the simple, easily calculated value
$\bbc$ might provide some insight into the number of planets that can exist
between two known planets. It would be interesting, then, to see if
this one
parameter can be used to test or  not whether a system is packed.

\acknowledgements

R. K gratefully acknowledges the support of National Science Foundation Grant
No.~PHY 06-53462 and No.~PHY 05-55615, and NASA Grant No.~NNG05GF71G, awarded
to The Pennsylvania State University.  S.N.R. and R.B. acknowledge funding
from NASA Astrobiology Institutes's Virtual Planetary Laboratory lead team,
supported by NASA under Cooperative Agreement No. NNH05ZDA001C.
The authors acknowledge the Research Computing and Cyberinfrastructure unit of
Information Technology Services at The Pennsylvania State University for
providing HPC resources and services that have contributed to the research
results reported in this paper. URL: http://rcc.its.psu.edu

\clearpage
\thispagestyle{empty}
\clearpage

\clearpage
\thispagestyle{empty}
\clearpage

\clearpage
\begin{deluxetable}{cccc}
\tablecaption{Orbital parameters of planets around HD 47186 
\citep{Bouchyetal2008}} 
\tablewidth{0pt}
\tablehead{ \colhead{Parameters} & \colhead{HD 47186b} & \colhead{HD 47186c}}
\startdata P [days] & 4.0845 $\pm$ 0.0002 & 1353.6 $\pm$ 57.1\\
e & 0.038 $\pm$ 0.020 & 0.249 $\pm$ 0.073 \\
msin $i$ ($M_\mathrm{Jup}$)&0.07167  & 0.35061 \\
$a$ (AU) & 0.050 & 2.395 \\
 \\
\enddata
\label{table1}
\end{deluxetable}
\clearpage

\end{document}